\documentclass[10pt]{article}
\usepackage{hyperref}
\usepackage{geometry}
\usepackage{graphicx}
\usepackage{slashed}
\usepackage{cite}
\usepackage{bbm}
\usepackage{dsfont}
\usepackage{color}
\usepackage{indentfirst}
\usepackage{multirow}
\usepackage{subfigure}
\usepackage{float}
\usepackage{makecell}
\usepackage{amsmath,amsfonts,amssymb}
\usepackage{tikz,pgf}
\usepackage{cleveref}
\usepackage{subcaption}
\usepackage{booktabs}
\usepackage[table]{xcolor} 

\usetikzlibrary{calc}
\usetikzlibrary{intersections}
\usetikzlibrary{trees}
\usetikzlibrary{decorations.pathmorphing}
\usetikzlibrary{decorations.markings}
\usetikzlibrary{patterns}
\tikzset{
   global scale/.style={
      scale=#1,
      every node/.append style={scale=#1}},
   photon/.style={decorate, decoration={snake}, draw=red},
   nucleon/.style={draw=black, postaction={decorate},
      decoration={markings,mark=at position .55 with{\arrow[draw=black]{>}}}},
   delta/.style={
      line width=2pt,
      double distance=1.5pt,
      draw=black,
      postaction={
        decorate},
        decoration={
          markings,
          mark=at position .55 with {\arrow[draw=black, scale=0.5]{>}}
        }
    },
   pion/.style={draw=blue, postaction={decorate},
      decoration={markings,mark=at position .55 with{\arrow[draw=blue]{}}}},
    }

\allowdisplaybreaks[4]

\begin{document}

\title{Studies of low energy $l+p\to l+p+\gamma$ process in covariant chiral perturbation theory }

\author{Xu Wang \textsuperscript{1,2}\footnote{wangxu0604@stu.pku.edu.cn},
Kai-Ge Kang \textsuperscript{2}\footnote{kaige@alumni.pku.edu.cn},
Zhiguang Xiao\textsuperscript{1}\footnote{xiaozg@scu.edu.cn (corresponding author)}
and 
Han-Qing Zheng \textsuperscript{1}\footnote{zhenghq@scu.edu.cn (corresponding author)}\\
\small \textsuperscript{1} Institute of Particle and Nuclear Physics, College of Physics, Sichuan University, Chengdu, Sichuan 610065, China\\
\small \textsuperscript{2} School of Physics, Peking University, Beijing 100871, China\\
}
\maketitle

\begin{abstract}
This study presents a tree-level calculation of the scattering amplitude for
the $lp\to lp\gamma$ (with a hard photon) process within the framework of
Chiral Perturbation Theory. Our calculations, based on the $O(p^2)$ and
$O(p^3)$ nucleon-pion Lagrangians, aim to provide a theoretical prediction for
the differential cross-section. The result shows that explicit inclusion of the
nonzero lepton mass significantly influences the low energy differential cross
section for $\mu p\to \mu p \gamma$ process. The kinematic region of the
present experimental data is beyond the validity domain of the $\chi$PT and is
therefore not suitable for determining the low-energy constants (LECs). By
comparing our results with future experimental data, we expect to determine the
values of the LECs as a further test of $\chi$PT as an effective low-energy
theory of QCD. The process is of significant interest as it can help to
determine the generalized polarizabilities of the nucleon.
\end{abstract}

\section{Introduction}

Quantum Chromodynamics (QCD), the fundamental theory describing strong interactions, faces significant computational challenges in the low-energy regime due to non-perturbative effects. On the other hand, Chiral Perturbation Theory ($\chi$PT), as a low-energy effective field theory of QCD, provides a rigorous framework to address these difficulties by preserving the essential features of chiral symmetry and its spontaneous breaking. Covariant Chiral Perturbation Theory, in particular, offers a systematic description of low-energy hadronic processes while maintaining Lorentz covariance \cite{Gasser:1983yg,Becher:1999he}. Most notably, the Extended-On-Mass-Shell (EOMS) scheme \cite{PhysRevD.68.056005},   
exhibits superior convergence properties compared to Heavy Baryon Chiral Perturbation Theory, providing a high-precision theoretical tool for studying low energy processes involving nucleons.\cite{Lensky_2009}

The success of covariant $\chi$PT has been verified across several critical domains. Beside baryon static properties  \cite{PhysRevLett.101.222002, Martin_Camalich_2010}, in $\pi N$ scatterings, $\text{B}\chi\text{PT}$ has laid the foundation for studying nucleon structure through the precise description of pion-proton interactions at low energies \cite{Chen_2013,Yao:2016vbz,Alarc_n_2013,Wang:2017agd}. In photoproduction processes, partial wave analysis based on covariant chiral theory has successfully elucidated the primary physical mechanisms underlying pion weak and photo productions~\cite{GuerreroNavarro:2019fqb,Yao:2018pzc,Ma:2020hpe,Cao:2021kvs}, $\pi^+\pi^-$ photon production \cite{Kang:2024fsf,GomezTejedor1996,Battaglieri2009},  etc.
These achievements provide a reliable theoretical benchmark for testing the Standard Model and exploring potential new physics effects.

Electron-proton ($ep$) elastic scattering remains a primary tool for probing the electromagnetic structure of the nucleon, supported by high-precision experimental data from   BINP, CERN, DESY, Fermilab, JLab, MAMI, and SLAC. These data serve as the basis for extracting the proton's electromagnetic form factors $G_E(Q^2)$ and $G_M(Q^2)$. Recently, interest in this field has intensified due to two major controversies: first, the discrepancy between $G_E/G_M$ ratios measured via polarization transfer versus the traditional Rosenbluth separation \cite{PERDRISAT2007694}; and second, the ``proton radius puzzle''---the unresolved inconsistency between the proton charge radius derived from muonic hydrogen Lamb shift experiments \cite{muonshift} and those from $ep$ scattering or electronic hydrogen spectroscopy \cite{RevModPhys.80.633,PhysRevLett.105.242001,annurev-nucl-102212-170627}. These ``puzzles'' have prompted new high-precision experiments at very low momentum transfer ($Q^2$), such as the PRad experiment at JLab \cite{Xiong:2019Nature}, COMPASS++/AMBER at CERN \cite{adams2019letterintentnewqcd,xiong2023protonchargeradiuslepton}, and MUSE at PSI \cite{gilman2013studyingprotonradiuspuzzle,gilman2017technicaldesignreportpaul}.

These experiments focus on the extremely low $Q^2$ region, placing stringent demands on theoretical accuracy. For instance, the MUSE experiment aims to measure $\mu p$ scattering cross-sections within $|Q^2| \sim 0.0016$--$0.08\ \text{GeV}^2$ to reach sub-percent precision on the charge radius. At these scales, traditional radiative correction methods---such as the Mo-Tsai formula, the ultra-relativistic approximation ($E \gg m_e$) \cite{PhysRevC.62.054320}, or the Soft Photon Approximation (SPA) \cite{PhysRev.122.1898}---become inadequate. A central conflict arises between detector characteristics and traditional assumptions: the PRad experiment's Hybrid Calorimeter (HyCal) features a large acceptance and high energy resolution capable of detecting MeV-level photons \cite{Xiong:2019Nature}. In contrast, the SPA assumes photon energies being far below detector thresholds, often neglecting actual detector response functions in favor of a simple ``hard cutoff'' assumption. This leads to systematic biases in calculating real photon contributions. Furthermore, for $\mu p$ scattering, the muon mass ($m_\mu \approx 105.7\ \text{MeV}$) is non-negligible, rendering the ultra-relativistic approximation entirely invalid.

The precise description of real photon emission is one of the decisive factor in the accuracy of radiative corrections, particularly for the three-body final state process $ep \to ep\gamma$. This process encompasses the Bethe-Heitler (BH) mechanism, Virtual Compton Scattering (VCS), and their interference. The photon's angular and energy distributions, relative to detector acceptance, directly impact the reliability of extracting $G_E(Q^2)$, $G_M(Q^2)$, and the proton radius. Consequently, a systematic theoretical study of $ep \to ep\gamma$ is not only essential for refining radiative corrections at very low $Q^2$ but is also a prerequisite for resolving the proton structure puzzles. Only through an accurate description of differential cross-sections and phase space can real photon contributions be reliably separated from experimental data, ensuring the correct isolation of elastic scattering signals.

In this work, we first present the Lagrangians and Feynman rules required for the $ep \to ep\gamma$ process, in section 2. Based on these rules, we calculate the $\mathcal{O}(p^3)$ tree-level amplitudes for both the BH and VCS processes to obtain the total scattering amplitude. Utilizing data from JLab E00-110 experiment\cite{defurne2015e00}, we perform a fit of the low-energy constants (LECs) appearing in the covariant chiral framework. Finally, we calculate the differential cross section for $\mu p$ scattering (where lepton mass is significant) and future hard-photon scattering studies---and provide theoretical predictions using our fitted LECs alongside those calculated by Fuchs et al. \cite{guerrero2020threshold}.

\section{Theoretical formalism}

\subsection{Chiral Perturbation Theory}
In the process of calculating the process $ep \to ep\gamma$, we  use Lagrangians up to $\mathcal{O}(p^3)$: $\mathcal{L}_{\pi N}^{(2)}$ and $\mathcal{L}_{\pi N}^{(3)}$.
The Lagrangian up to $O(p^3)$ \cite{fettes2000chiral} reads:
\begin{align}
      \begin{aligned}
      \mathcal{L}_{\pi N}^{(1)}&=\bar{\Psi}\Bigl(
         i\slashed{D}-m+\frac{g_A}{2}\slashed{u}\gamma_5
      \Bigr)\Psi,\\
      \mathcal{L}_{\pi N}^{(2)}&=c_1 \langle \chi_+\rangle + \bar{\Psi}\sigma^{\mu\nu}[\frac{c_6}{8m}F_{\mu\nu}^+ + \frac{c_7}{8m}\langle F_{\mu\nu}^{+}\rangle]\Psi,\footnotemark
      \\
      \mathcal{L}_{\pi N}^{(3)}&= \bar{\Psi} \Bigl(\frac{d_6}{2 m} i\left[D^\mu, \widetilde{F}_{\mu \nu}^{+}\right] D^\nu+ \text{h.c.}\Bigr) \Psi + \bar{\Psi} \Bigl(\frac{d_7}{2 m} i\left[D^\mu,\left\langle F_{\mu \nu}^{+}\right\rangle\right] D^\nu+ \text{h.c.}\Bigr) \Psi,
      \end{aligned}
\end{align}
\footnotetext{The definitions of the $c_6$ and $c_7$ terms used here differ slightly from those in Scherer's work \cite{scherer2005}. The relationship between the two sets of coefficients is $c_6=4m_N c_6^{\prime}$, $c_7=m_N(-2c_6^{\prime}+c_7^{\prime})$.\label{lecs_footnote}}
where terms are defined as follows: $D_{\mu}\Psi=\Bigl(\partial_{\mu}+\Gamma_{\mu}\Bigr)\Psi$, and
\begin{align}
   \begin{aligned}
  \Gamma_{\mu}&=\frac{1}{2}[u^{\dagger}(\partial_{\mu}-ir_{\mu})u+u(\partial_{\mu}-il_{\mu})u^{\dagger}],\\
u^{\mu}&=i[u^{\dagger}(\partial_{\mu}-ir_{\mu})u-u(\partial_{\mu}-il_{\mu})u^{\dagger}],\\
\chi_{\pm}&=u^{\dagger}\chi u^{\dagger}\pm u\chi^{\pm}u,\\
F_{\mu\nu}^{\pm}&=uF_{L\mu\nu}u^{\dagger}\pm u^{\dagger}F_{R\mu\nu}u,\\
F_{L,\mu\nu}&=\partial_{\mu}l_{\nu}-\partial_{\nu}l_{\mu}-i[l_{\mu},l_{\nu}],\\
F_{R,\mu\nu}&=\partial_{\mu}r_{\nu}-\partial_{\nu}r_{\mu}-i[r_{\mu},r_{\nu}].
\end{aligned}
\end{align}
Here, the symbols denote the following: $\langle X \rangle$ represents the trace, $\tilde{X} = X - \frac{1}{2} \langle X \rangle$.

Some common terms in the Lagrangian, after expanding the meson fields, are as follows:
\begin{align}
   \begin{aligned}
      &u = \mathbbm{1} + i\frac{\phi}{2F} -\frac{\phi^2}{8F^2} + \mathcal{O}(\phi^3),\\
      &\Gamma_{\mu} = -ir_{\mu} + \frac{1}{8F^2}[\phi \partial_{\mu}\phi - (\partial_{\mu}\phi)\phi] - \frac{i}{4F^2}\phi r_{\mu}\phi +\frac{i}{8F^2}\phi^2 r_{\mu} + \frac{i}{8F^2}r_{\mu}\phi^2 + \mathcal{O}(\phi^3),
      \\
      &u_{\mu} = -\frac{1}{F}\partial_{\mu}\phi + \frac{i}{F}r_{\mu}\phi -\frac{i}{F}\phi r_{\mu} + \mathcal{O}(\phi^3),\\
      &\chi_{+} = 2M^2\mathbbm{1} - \frac{1}{F^2}M^2\phi^2 + \mathcal{O}(\phi^3),\\
      &\chi_{-}=-\frac{2i}{F}M^2\phi^2 + \mathcal{O}(\phi^3),\\
      &F_{L,\mu\nu}=F_{R,\mu\nu}=-eF_{\mu\nu}Z,\\
      &F_{\mu\nu}^{+} = 2F_{R,\mu\nu} + \frac{1}{2F^2}\phi F_{R,\mu\nu}\phi -\frac{1}{4F^2}\phi^2F_{R,\mu\nu}-\frac{1}{4F^2}F_{R,\mu\nu}\phi^2 + \mathcal{O}(\phi^3),\\
      &F_{\mu\nu}^{-} = \frac{i}{F}(\phi F_{R,\mu\nu}-F_{R,\mu\nu}\phi) + \mathcal{O}(\phi^3).
   \end{aligned}
\end{align}

When introducing the photon field, we take
\begin{align}
r_{\mu} = l_{\mu} = -eA_{\mu}Z,\quad Z=\frac{\mathbbm{1}+\tau_3}{2}.
\end{align}

Then we can obtain the following Feynman rules of the interaction vertex in Figure~\ref{fig:3-vertex}:
\begin{figure}[htbp]
   \centering
   \begin{tikzpicture}[scale=1]
       \draw[photon](0,0)to(0,-2);
       \draw[nucleon, thick](-2,-2)to(0,-2);
       \draw[nucleon, thick](0,-2)to(2,-2);            
       \node[below] at (-1,-2) {$p_1$};
       \node[below] at (1,-2) {$p_2$};
       \node[right] at (0.2,-1) {$l_1$};
       \draw[nucleon, thick](0.2,-0.5)to(0.2,-1.5);
       \draw[fill] (0,-2) circle (0.1);
   \end{tikzpicture}
   \caption{The $\gamma p$ interaction vertex for $e p\to e p\gamma$ up to $O(p^3)$.\label{fig:3-vertex}}
\end{figure}
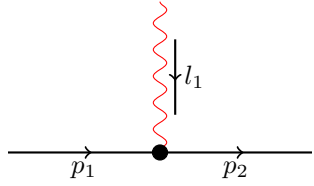

\begin{align}
   \begin{aligned}
       O(p^1):& -ie\gamma^{\mu}, \\
       O(p^2):& -ie (c_6+c_7)\gamma^{\mu} + \frac{ie(c_6+c_7)}{2m_N}(p_1+p_2)^{\mu},\\
       O(p^3):& \frac{ie(d_6+2d_7)(p_1-p_2)^2}{m_N}(p_1+p_2)^{\mu}.
       \label{feyrules}
   \end{aligned}
\end{align}

\subsection{Feynman Diagram Calculations}

\begin{figure}[H]
    \centering
    \subfigure[]{
    \begin{minipage}[t]{0.3\textwidth}
        \centering
        \begin{tikzpicture}[scale=0.5]
            \draw[nucleon](-3,1.5)to(0,0);
            \draw[nucleon](0,0)to(3,1.5);
            \draw[nucleon, very thick](-3,-4.5)to(0,-3);
            \draw[nucleon, very thick](0,-3)to(3,-4.5);
            \draw[photon](0,0)to(0,-3);
            \draw[photon](-2,1)to(-1,3);
            \draw[fill] (0,-3) circle (0.3);
            \node[above] at (-1.8,1.8) {$k$};
            \node[below] at (-2.5,1) {$k_1$};
            \node[below] at (2.5,1) {$k_2$};
            \node[above] at (-2.5,-4) {$p_1$};
            \node[above] at (2.5,-4) {$p_2$};
            
        \end{tikzpicture}
        \label{fig2.eii}
    \end{minipage}}
    \subfigure[]{
    \begin{minipage}[t]{0.3\textwidth}
        \centering
        \begin{tikzpicture}[scale=0.5]
            \draw[nucleon](-3,1.5)to(0,0);
            \draw[nucleon](0,0)to(3,1.5);
            \draw[nucleon, very thick](-3,-4.5)to(0,-3);
            \draw[nucleon, very thick](0,-3)to(3,-4.5);
            \draw[photon](2,1)to(3,3);
            \draw[photon](0,0)to(0,-3);
            \draw[fill] (0,-3) circle (0.3);
            \node[above] at (2,1.8) {$k$};
            \node[below] at (-2.5,1) {$k_1$};
            \node[below] at (2.5,1) {$k_2$};
            \node[above] at (-2.5,-4) {$p_1$};
            \node[above] at (2.5,-4) {$p_2$};
            \label{fig2.eff}
        \end{tikzpicture}
    \end{minipage}}
    \centering
    \subfigure[]{
    \begin{minipage}[t]{0.3\textwidth}
        \centering
        \begin{tikzpicture}[scale=0.5]
            \draw[nucleon](-3,1.5)to(0,0);
            \draw[nucleon](0,0)to(3,1.5);
            \draw[nucleon, very thick](-3,-4.5)to(0,-3);
            \draw[nucleon, very thick](0,-3)to(3,-4.5);
            \draw[photon](0,0)to(0,-3);
            \draw[photon](0,-3)to(3,-2);
            \draw[fill] (0,-3) circle (0.5);
            \node[above] at (3,-2) {$k$};
            \node[below] at (-2.5,1) {$k_1$};
            \node[below] at (2.5,1) {$k_2$};
            \node[above] at (-2.5,-4) {$p_1$};
            \node[above] at (2.5,-4) {$p_2$};

        \end{tikzpicture}
        \label{fig2.p}
    \end{minipage}}
    \caption{The $e p\to e p\gamma$ process.}
    \label{fig2.fushe}
\end{figure}
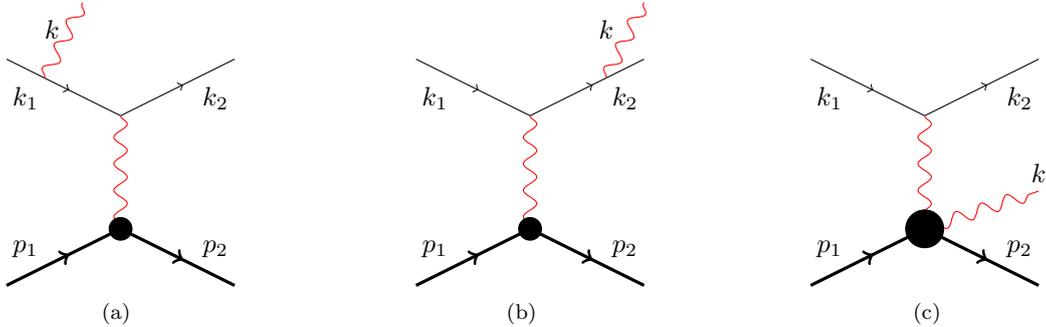

The Feynman diagrams for the $ep\to ep\gamma$ process can be broadly divided into two categories: a photon radiating from the lepton (electron) leg, and a photon radiating from the hadron (proton) leg. These two categories can be further subdivided into three distinct diagram structures, as shown in Figure~\ref{fig2.fushe}.

Figures \ref{fig2.eii} and \ref{fig2.eff} depict the electron bremsstrahlung process, whose amplitudes are primarily described by QED theory, while the proton part is simplified to a ``black box'' described by form factors. Figure \ref{fig2.p} describes a virtual Compton scattering process, where a virtual photon interacts with a proton and radiates a real photon. The amplitudes of these three diagrams must be calculated separately and summed to obtain the total scattering amplitude.

For diagrams where the photon radiates from the lepton leg, the amplitude can be decomposed into a leptonic part and a hadronic part. The amplitude $\mathcal{M}_a, \mathcal{M}_b$ for Figure \ref{fig2.eii} and \ref{fig2.eff}, where the sum $\mathcal{M}_a + \mathcal{M}_b$ represents the BH process, can be written as: 
\begin{align}
    \begin{aligned}
    \mathcal{M}_{a}&=\bar{u}(k_2,m_e)(-ie\gamma^{\mu})\frac{i(\slashed{k}_1-\slashed{k}+m_e)}{(k_1-k)^2-m_e^2}(-ie\gamma^{\nu})u(k_1,m_e)\frac{-i}{(p_1-p_2)^2}\bar{u}(p_2,m_N)\Gamma_{\mu}u(p_1,m_N)\epsilon_{\nu}^{*}\\
    &=\frac{e^2}{(p_1-p_2)^2}\frac{-1}{(k_1-k)^2-m_e^2}\bar{u}(k_1,m_e)\gamma^{\mu}(\slashed{k}_1-\slashed{k}+m_e)\gamma^{\nu}u(k_1,m_e)\bar{u}(p_2,m_N)\Gamma_{\mu}u(p_2,m_N)\epsilon_{\nu}^{*},\\
     \mathcal{M}_{b}&=\bar{u}(k_2,m_e)(-ie\gamma^{\nu})\frac{i(\slashed{k}+\slashed{k}_2+m_e)}{(k+k_2)^2-m_e^2}(-ie\gamma^{\mu})u(k_1,m_e)\frac{-i}{(p_1-p_2)^2}\bar{u}(p_1,m_N)\Gamma_{\mu}u(p_2,m_N)\epsilon_{\nu}^{*}\\
        &=\frac{e^2}{(p_1-p_2)^2}\frac{-1}{(k+k_2)^2-m_e^2}\bar{u}(k_1,m_e)\gamma^{\nu}(\slashed{k}+\slashed{k}_2+m_e)\gamma^{\mu}u(k_1,m_e)\bar{u}(p_1,m_N)\Gamma_{\mu}u(p_1,m_N)\epsilon_{\nu}^{*}.
    \end{aligned}
\end{align}

In the expressions above, the hadronic part is described by $\Gamma_{\mu}$, which contains nucleon form factors. It can generally be decomposed into the Dirac form factor $F_1^N(Q_1^2)$ and the Pauli form factor $F_2^N(Q_1^2)$: 
\begin{align}
    \Gamma^{\mu}=F_1^N(Q_1^2)\gamma^{\mu}+i\frac{\sigma^{\mu\nu}q_{\nu}}{2m_N}F_2^N(Q_1^2).
\end{align}
where $q_1^\mu=p^\mu_2-p^\mu_1$ and $Q_1^2=-q_1^2$. These form factors summarize the influence of the proton's internal structure on electromagnetic interactions, rather than behaving as a simple point particle.

For the virtual Compton scattering process in Figure \ref{fig2.p}, its amplitude $\mathcal{M}_{c}$ can be written in the following form:
\begin{align}
    \begin{aligned}
        \mathcal{M}_{c}&=\bar{u}(k_2,m_e)(-ie\gamma_{\mu})u(k_1,m_e)\frac{-i}{l_1^2}\bar{u}(p_2,m_N)\Gamma^{\mu\nu}u(p_1,m_N)\epsilon_{\nu}^{*}\\
        &=\frac{-1}{l_1^2}e\bar{u}(k_2,m_e)\gamma_{\mu}u(k_1,m_e)\bar{u}(p_2,m_N)\Gamma^{\mu\nu}u(p_1,m_N)\epsilon_{\nu}^{*}.
    \end{aligned}
\end{align}
where $\Gamma^{\mu\nu}$  represents the hadronic part where a virtual photon interacts with the proton and radiates a real photon. For simplification, this hadronic part can be rearranged and defined as the virtual Compton scattering amplitude $M_{c}^{\mu\nu}$, with the expression $\overline{u}(p_2,m_N)(\Gamma^{\mu\nu})u(p_1,m_N)\epsilon_{\nu}^{*}\equiv M_{c}^{\mu\nu}\epsilon_{\nu}^{*}$.
For ease of calculation, the hadronic amplitude $M_{c}^{\mu\nu}$ is usually decomposed according to its Lorentz structure :
$$M_{c}^{\mu\nu}=\sum_{i}\bar{u}(p_2,m_N)A_{i}\Gamma_{i}^{\mu\nu}u(p_1,m_N),$$
{where $\Gamma_{i}^{\mu\nu}$ is a combination of 26 fundamental Lorentz structures composed of $g^{\mu\nu}$, $\gamma^\mu$, $k^\mu$, $p^\mu_1$, $p^\mu_2$ \footnotemark,}\footnotetext{The 34 Lorentz structures for double virtual photons scattering identified by \cite{tarrach1975invariant} are reduced to 26 under the constraint $k \cdot \epsilon^* = 0$. We have carefully checked that the amplitude as presented by Table~\ref{table_amp} satifies gauge invariance. } such as $g^{\mu\nu}$, $\gamma^{\mu}\gamma^{\nu}$,  $\gamma^{\mu}\gamma^{\nu}\slashed k$, as shown in Table \ref{table_amp}. This decomposition is a common method for handling complex hadronic amplitudes, transforming the problem of calculating the amplitude into solving for a series of scalar coefficients $A_i$.

For virtual Compton scattering, the Feyman diagrams at tree level up to $O(p^3)$ are shown in Figure~\ref{vcsdiagram}.
Let the four-momentum of the virtual photon be $l_1 = k + p_2 - p_1$, we obtain the following results:
\begin{figure}[H]
    \centering
    \begin{minipage}[t]{0.3\textwidth}
        \centering
        \begin{tikzpicture}[scale=0.5]
            \draw[photon](-3,2)to(-1,0);
            \draw[nucleon, thick](-1,0)to(1,0);
            \draw[photon](1,0)to(3,2);
            \draw[nucleon, thick](-3,-2)to(-1,0);
            \draw[nucleon, thick](1,0)to(3,-2);
            \node[below] at(-2.2,-1.5) {$p_1$};
            \node[below] at(2.2,-1.5) {$p_2$};
            \node[above] at(-2.2,1.5) {$l_1$};
            \node[above] at(2.2,1.5) {$k$};
            \fill (-1,0) circle (5pt);
            \fill (1,0) circle (5pt);
        \end{tikzpicture}
        \caption*{(1)}
    \end{minipage}
    \begin{minipage}[t]{0.3\textwidth}
        \centering
        \begin{tikzpicture}[scale=0.8]
            \draw[photon](-2,0.5) to(0,0);
            \draw[nucleon, thick](-2,-2.5)to(0,-2);
            \fill (0,0) circle (3pt);
            \fill (0,-2) circle (3pt);
            \draw[nucleon, thick](0,-2) to (0,0);
            \draw[nucleon, thick](0,0) to (2,-2.5);
            \draw[photon](0,-2) to (2,0.5);
            \node[above] at(-1.5,-2.2) {$p_1$};
            \node[below] at(-1.5,1.2) {$l_1$};
            \node[below] at(1.5,1.0) {$k$};
            \node[above] at(1.8,-2) {$p_2$};
        \end{tikzpicture}
        \caption*{(2)}
    \end{minipage}
    \begin{minipage}[t]{0.3\textwidth}
        \centering
        \begin{tikzpicture}[scale=0.7]
            \draw[photon](-2,2) to(0,0);
            \draw[nucleon, thick](-2,-2)to(0,0);
            \fill (0,0) circle (3pt);
            \draw[nucleon, thick](0,0) to (2,-2);
            \draw[photon](0,0) to (2,2);
            \node[below] at(-1.2,-1.5) {$p_1$};
            \node[below] at(1.2,-1.5) {$p_2$};
            \node[below] at(-1.5,2.4) {$l_1$};
            \node[below] at(1.5,2.4) {$k$};
        \end{tikzpicture}
        \caption*{(3)}
    \end{minipage}
    \caption{Virtual Compton Scattering diagrams.}
    \label{vcsdiagram}
\end{figure}

\begin{align}
    \begin{aligned}
       {  \mathcal{M}_1}=&i\bar{u}(p_2,m_N)\cdot\Bigl(
            \frac{ie(c_6+c_7)\slashed{k}\slashed{\epsilon}^*(k)}{4m_N} - \frac{ie(c_6+c_7)\slashed{\epsilon}^*(k) \slashed{k}}{4m_N} +\frac{ie(d_6+2d_7)k^2 p_2\cdot \epsilon^*(k)}{m_N} -ie \slashed{\epsilon}^*(k)
         \Bigr)\\
         &\frac{(\slashed{p}_2+\slashed{k})+m}{(k+p_2)^2-m^2}\cdot
        \Bigl(
           -\frac{ie (c_6+c_7)\slashed{l}_1 \slashed{\epsilon}(l_1)}{4m_N} +\frac{ie(c_6+c_7)\slashed{\epsilon}(l_1)\slashed{l}_1}{4m_N} - ie \slashed{\epsilon}(l_1)+ \\
           &\frac{ie(d_6+2d_7) l_1 \cdot \epsilon(l_1)(- l_1 \cdot(k+p_2+p_1))}{2m_N} -\frac{ie(d_6+2d_7)l_1^2(-\epsilon(l_1)\cdot(k+p_2+p_1))}{2m_N}
        \Bigr)u(p_1,m_N),
    \end{aligned}
\end{align}
\begin{align}
    \begin{aligned}
        \mathcal{M}_2=&i\bar{u}(p_2,m_N)\Bigl(
           -\frac{ie (c_6+c_7)\slashed{l}_1 \slashed{\epsilon}(l_1)}{4m_N} +\frac{ie(c_6+c_7)\slashed{\epsilon}(l_1)\slashed{l}_1}{4m_N} - ie \slashed{\epsilon}(l_1)\\
           &-\frac{ie(d_6+2d_7)l_1\cdot \epsilon(l_1)(l_1\cdot(p_1+p_2-k))}{2m_N} +\frac{ie(d_6+2d_7)l_1^2(\epsilon(l_1)\cdot(p_1+p_2-k))}{2m_N}
        \Bigr)\cdot
         \frac{(\slashed{p}_1-\slashed{k})+m}{(k-p_1)^2-m^2}\cdot\\
        &\Bigl(
           \frac{ie(c_6+c_7)\slashed{k}\slashed{\epsilon}^*(k)}{4m_N} - \frac{ie(c_6+c_7)\slashed{\epsilon}^*(k) \slashed{k}}{4m_N} +\frac{ie(d_6+d_7)k^2(p_1\cdot \epsilon^*(k))}{m_N}-ie \slashed{\epsilon}^*(k)
        \Bigr)u(p_1,m_N),
    \end{aligned}
\end{align}

\begin{align}
    \begin{aligned}
        \mathcal{M}_3=\bar{u}(p_2,m_N)  \frac{i e^2 (d_6+2d_7)}{m_N}([l_1\cdot \epsilon(l_1)][l_1\cdot\epsilon^*(k)]-k^2[\epsilon(l_1)\cdot \epsilon^*(k)]-l_1^2[\epsilon(l_1)\cdot \epsilon^*(k)])u(p_1,m_N),
    \end{aligned}
\end{align}
{where $\mathcal{M}_{1,2,3}\equiv \mathcal{M}^{{(1,2,3)}\mu\nu}\epsilon_\mu(l_1)\epsilon_\nu^*(k)$ represent the individual contributions to the virtual Compton amplitudes as shown in Figure~\ref{vcsdiagram}. The total amplitude is given by the sum $\mathcal{M}_{\text{vcs}} = \mathcal{M}_1 + \mathcal{M}_2 + \mathcal{M}_3 = \epsilon_{\mu}(l_1) \epsilon_{\nu}^*(k) M_{c}^{\mu\nu}$. When coupling this to the leptonic part, one simply needs to replace the photon polarization vector $\epsilon_{\mu}(l_1)$ with the product of the photon propagator and the corresponding leptonic current: $\frac{-1}{l_1^2} e \bar{u}(k_2, m_e) \gamma_{\mu} u(k_1, m_e)$. Notice that some expressions given above  vanish
with this replacement owing to gauge invariance.
The expressions given above incorporate higher-order terms that require truncation in accordance with the chiral power counting scheme outlined in \cite{scherer2005}. Given that the vertices are of order \(O(p^3)\) at most, the resulting amplitude can extend up to \(O(p^5)\). On the other side, since the amplitude is calculated up to \(O(p^3)\) in this work, its squared magnitude consequently reaches up to \(O(p^6)\). To ensure theoretical consistency throughout the calculation, we neglect all terms in the amplitude beyond $\mathcal{O}(p^3)$. Note that   terms appearing in \(|\mathcal{M}|^2\) at orders higher than \(O(p^3)\) are not  removed.

We   define the five Mandelstem variables as follows:
\begin{align}
    \begin{aligned}
        s=(k_1+&p_1)^2,\quad s_1=(k+p_2)^2,\quad s_2=(k+k_2)^2,\\
        &t_1=(k-p_1)^2,\quad t_2=(p_1-p_2)^2.
    \end{aligned}
\end{align}
Based on the above definition, the scattering amplitude can be systematically simplified. After truncation and rearrangement, the results for each order of the amplitude are shown in Table~\ref{table_amp}.
 
\begin{table}[H]
    \centering
    \resizebox{\linewidth}{!}{
        \renewcommand{\arraystretch}{2} 
        {\Huge
    \begin{tabular}{ccccc}
      \hline
      \rowcolor{green!40}
            & $\gamma^{\mu}\gamma^{\nu}$ & $\gamma^{\mu}\gamma^{\nu}\slashed{k}$ & $g^{\mu\nu}$ &$g^{\mu\nu}\slashed{k}$\\
      \hline
      $\mathcal{M}^{(1)\mu\nu}$& $0$ & $\frac{ie^2}{m_N^2-s_1}+\frac{ie^2}{m_N^2-t_1}$ & $0$ &$-\frac{-2ie^2}{m_N^2-s_1}$\\
      \hline
      $\mathcal{M}^{(2)\mu\nu}$& 0 & $\frac{2m_N A}{m_N^2-s_1}+\frac{2A}{m_N^2-t_1}$& $-2A$ &$-\frac{4A}{m_N^2-s_1}$\\
      \hline
      $\mathcal{M}^{(3)\mu\nu}$& 0&$\frac{B(3m_N^2+s_1)}{4(m_N^2-s_1)}+\frac{B(3m_N^2+t_1)}{4(m_N^2-t_1)}$ &$\frac{C(2m_N^2-s_1-t_1-t_2)}{m_N}-B$ &$-\frac{B(3m_N^2+s_1)}{2(m_N^3-m_N t_1)}$\\
      \hline
      \rowcolor{green!40}
      & $k^{\mu}\gamma^{\nu}$ & $k^{\mu}\gamma^{\nu}\slashed{k}$ & $p_1^{\mu}\gamma^{\nu}$ &$p_1^{\mu}\gamma^{\nu}\slashed{k}$\\
      \hline
      $\mathcal{M}^{(1)\mu\nu}$& $\frac{2ie^2}{m_N^2-s_1}$ & 0 & 0 &0\\
      \hline
      $\mathcal{M}^{(2)\mu\nu}$& $\frac{4m_N A}{m_N^2-s_1}$ & $-\frac{A}{2(m_N^2-s_1)}+\frac{A}{2(m_N^2-t_1)}$ & 0 &$-\frac{A}{2(m_N^2-s_1)}-\frac{A}{2(m_N^2-t_1)}$\\
      \hline
      $\mathcal{M}^{(3)\mu\nu}$& $-\frac{B(7m_N^2+s_1)}{4(m_N^2-s_1)}-\frac{1}{4}B$& $\frac{C(m_N^2-t_1-t_2)-Bm_N^2}{2(m_N^3-m_N s_1)}+\frac{Bm_N^2-C(m_N^2- s_1-t_2)}{2(m_N^3-m_N t_1)}$&$\frac{1}{2}B$ &$\frac{C(3m_N^2-2s_1-t_1-t_2)-Bm_N^2}{2(m_N^3-m_N s_1)}+\frac{C(m_N^2-s_1-t_2)-Bm_N^2}{2(m_N^3-m_N t_1)}$\\
      \hline
      \rowcolor{green!40}
      & $p_2^{\mu} \gamma^{\nu}$ & $p_2^{\mu} \gamma^{\nu}\slashed{k}$ & $p_1^{\nu}\gamma^{\mu}$ &$p_1^{\nu}\gamma^{\mu}\slashed{k}$\\
      \hline
      $\mathcal{M}^{(1)\mu\nu}$& 0& 0 & $\frac{2ie^2}{m_N^2-t_1}$ &0\\
      \hline
      $\mathcal{M}^{(2)\mu\nu}$& 0 & $-\frac{A}{2(m_N^2-s_1)}-\frac{A}{2(m_N^2-t_1)}$ & $\frac{m_N A}{m_N^2-t_1}$&$\frac{A}{m_N^2-t_1}$\\
      \hline
      $\mathcal{M}^{(3)\mu\nu}$& $\frac{1}{2}B$ & $\frac{C(m_N^2-t_1-t_2)-Bm_N^2}{2(m_N^3-m_N s_1)}+\frac{C(3m_N^2-s_1-2t_1-t_2)-Bm_N^2}{(2m_N^3-m_N t_1)}$ & 0 &$\frac{C(m_N^2-t_1-t_2)}{m_N^3-m_N s_1}+\frac{B m_N}{m_N^2-t_1}$\\
      \hline
      \rowcolor{green!40}
      & $p_2^{\nu} \gamma^{\mu}$ & $p_2^{\nu}\gamma^{\mu}\slashed{k}$ & $p_1^{\mu}p_1^{\nu}$ &$p_1^{\mu}p_1^{\nu}\slashed{k}$\\
      \hline
      $\mathcal{M}^{(1)\mu\nu}$& $\frac{2ie^2}{m_N^2-s_1}$ & 0 & 0 &0\\
      \hline
      $\mathcal{M}^{(2)\mu\nu}$& $\frac{m_N A}{m_N^2-s_1}$ & $\frac{A}{m_N^2-s_1}$& $-\frac{A}{m_N^2-t_1}$ &0\\
      \hline
      $\mathcal{M}^{(3)\mu\nu}$&  0&  $\frac{B m_N}{m_N^2-s_1}$& $\frac{C}{m_N}+\frac{C(m_N^2-s_1-t_2)}{m_N^3-m_N t_1}$ &$\frac{C(m_N^2-t_1-t_2)}{m_N^3-m_N s_1}-\frac{B}{2(m_N^2-t_1)}$ \\
      \hline
      \rowcolor{green!40}
      & $p_2^{\mu}p_2^{\nu}$ & $p_2^{\mu}p_2^{\nu}\slashed{k}$ & $k^{\mu} p_1^{\nu}$ &$k^{\mu} p_1^{\nu} \slashed{k}$\\
      \hline
      $\mathcal{M}^{(1)\mu\nu}$& 0 & 0 & 0 &0\\
      \hline
      $\mathcal{M}^{(2)\mu\nu}$& $-\frac{A}{m_N^2-s_1}$ & 0 & $\frac{A}{m_N^2-t_1}$&0\\
      \hline
      $\mathcal{M}^{(3)\mu\nu}$& $\frac{C}{m_N}+\frac{C(m_N^2-t_1-t_2)}{m_N^3-m_N s_1}$ & $\frac{B}{2(m_N^2-s_1)}$ & $-\frac{C}{m_N}-\frac{C(m_N^2-s_1-t_2)}{m_N^3-m_N t_1}$ &$\frac{B}{2(m_N^2-t_1)}$\\
      \hline
      \rowcolor{green!40}
       & $k^{\mu}p_2^{\nu}$ &$k^{\mu}p_2^{\nu}\slashed{k}$& $p_1^{\mu}p_2^{\nu}$& $p_1^{\mu}p_2^{\nu}\slashed{k}$\\
      \hline
      $\mathcal{M}^{(1)\mu\nu}$& 0 & 0 & 0 &0\\
      \hline
      $\mathcal{M}^{(2)\mu\nu}$& $-\frac{3A}{m_N^2-s_1}$ & 0 & $-\frac{A}{m_N^2-s_1}$ &0\\
      \hline
      $\mathcal{M}^{(3)\mu\nu}$& $\frac{C}{m_N}+\frac{C(m_N^2-t_1-t_2)}{m_N^3-m_N s_1}-\frac{2 B m_N}{m_N^2-s_1}$ & $\frac{C}{2(m_N^2-s_1)}$ & $-\frac{C}{m_N}+\frac{C(3m_N^2-2s_1-t_1-t_2)}{m_N^3-m_N s_1}$ &$\frac{B}{2(m_N^2-s_1)}$\\
      \hline
      \rowcolor{green!40}
       & $p_2^{\mu}p_1^{\nu}$ &$p_2^{\mu}p_1^{\nu}\slashed{k}$&&\\
      \hline
      $\mathcal{M}^{(1)\mu\nu}$& 0 & 0 &  &\\
      \hline
      $\mathcal{M}^{(2)\mu\nu}$& $-\frac{A}{m_N^2-t_1}$ & 0 &  &\\
      \hline
      $\mathcal{M}^{(3)\mu\nu}$& $-\frac{C}{m_N}+\frac{C(3m_N^2-s_1-2t_1-t_2)}{m_N^3-m_N t_1}$ & $-\frac{B}{2(m_N^2-t_1)}$ &  &\\
      \hline
    \end{tabular}}
    }
    \caption{Virtual Compton Scattering amplitudes, where $A=ie^2(c_6+c_7)$, $B=ie^2(c_6+c_7)^2$, and $C=ie^2(d_6+2d_7)$.}
    \label{table_amp}
  \end{table}
  
  It is important to note a key detail here, which is the use of mass in the proton propagator. The $c_1$ term in the Lagrangian introduces a mass correction, i.e., $m\to m_2=m - 4c_1 M^2$ . However, $m_2-m_N \sim O(p^3)$, and this mass difference only appears in $O(p^3)$ loop diagram calculations. Since this is a tree diagram calculation, the propagator mass can be directly replaced with the true nucleon mass $m_N$. For example, the original form of the \( \gamma^{\mu} \gamma^{\nu} \slashed{k} \) term in \( M^{(1)\mu\nu} \) is \( \frac{ie^2}{m_2^2 - s_1} +\frac{ie^2}{m_2^2-t_1}\), but after applying the substitutions, it is expressed as shown in  Table~\ref{table_amp}.
  
\subsection{Kinematics and Formalism}

In the context of the $e(k_1)+p(p_1) \rightarrow e(k_2)+p(p_2)+\gamma(k)$ process, experimental data is typically expressed in terms of a set of standard variables, including the virtuality $Q^2$, the Bjorken scaling variable $x_B$, and the momentum transfer $t$. These are defined as follows \cite{defurne2015e00}:
\begin{align}
Q^2=-q^2=-(k_1-k_2)^2,\quad x_B=Q^2/(2q\cdot p_1),\quad t=(p_2-p_1)^2.
\end{align}
Additionally, the azimuthal angle $\phi$ is defined as the angle between the lepton plane and the hadron plane in the rest frame of the proton~\cite{Boer_1998,Mulders_1996} and can be expressed covariantly as~\cite{phiAngle}. 
\begin{align}
{\rm cos} \phi = -\frac{g^{\mu\nu}_{\bot}k_{1 \mu}p_{2\nu}}{|k_{1\bot}||p_{2\bot}|},  \quad
{\rm sin} \phi = -\frac{\varepsilon^{\mu\nu}_{\bot}k_{1\mu}p_{2\nu}}{|k_{1\bot}||p_{2\bot}|},
\label{cosphi}
\end{align}
where
\begin{align}
|k_{1\bot}|=&\sqrt{-g^{\mu\nu}_{\bot}k_{1\mu} k_{1\nu}},\quad |p_{2\bot}|=\sqrt{-g^{\mu\nu}_{\bot}p_{2\mu} p_{2\nu}},\\
g^{\mu\nu}_\bot =&g^{\mu\nu} - \frac{q^{\mu}p_1^{\nu}+p_1^{\mu}q^{\nu}}{p_1\cdot q(1+\gamma^2)} + \frac{\gamma^2}{1+\gamma^2}(\frac{q^{\mu}q^{\nu}}{-q^2}-\frac{p_1^\mu p_1^\nu}{m_N^2}),\\
\varepsilon^{\rho\sigma}_{\bot}=&\varepsilon^{\mu\nu\rho\sigma}\frac{p_{1\mu} q_\nu}{p_1\cdot q\sqrt{1+\gamma^2}}.
\end{align}
where $\gamma=2x_B m_N/Q,q=k_1-k_2$.
The schematic diagram of the $\phi$ angle is shown in Figure \ref{fig:axes}. The lepton plane is formed by the incident and scattered leptons, whereas the hadron plane is defined by the momentum transfer vector --- given by the difference between the incident and scattered lepton momenta --- and the outgoing proton.

\begin{figure}[htbp]
    \centering
    \includegraphics[width=\textwidth]{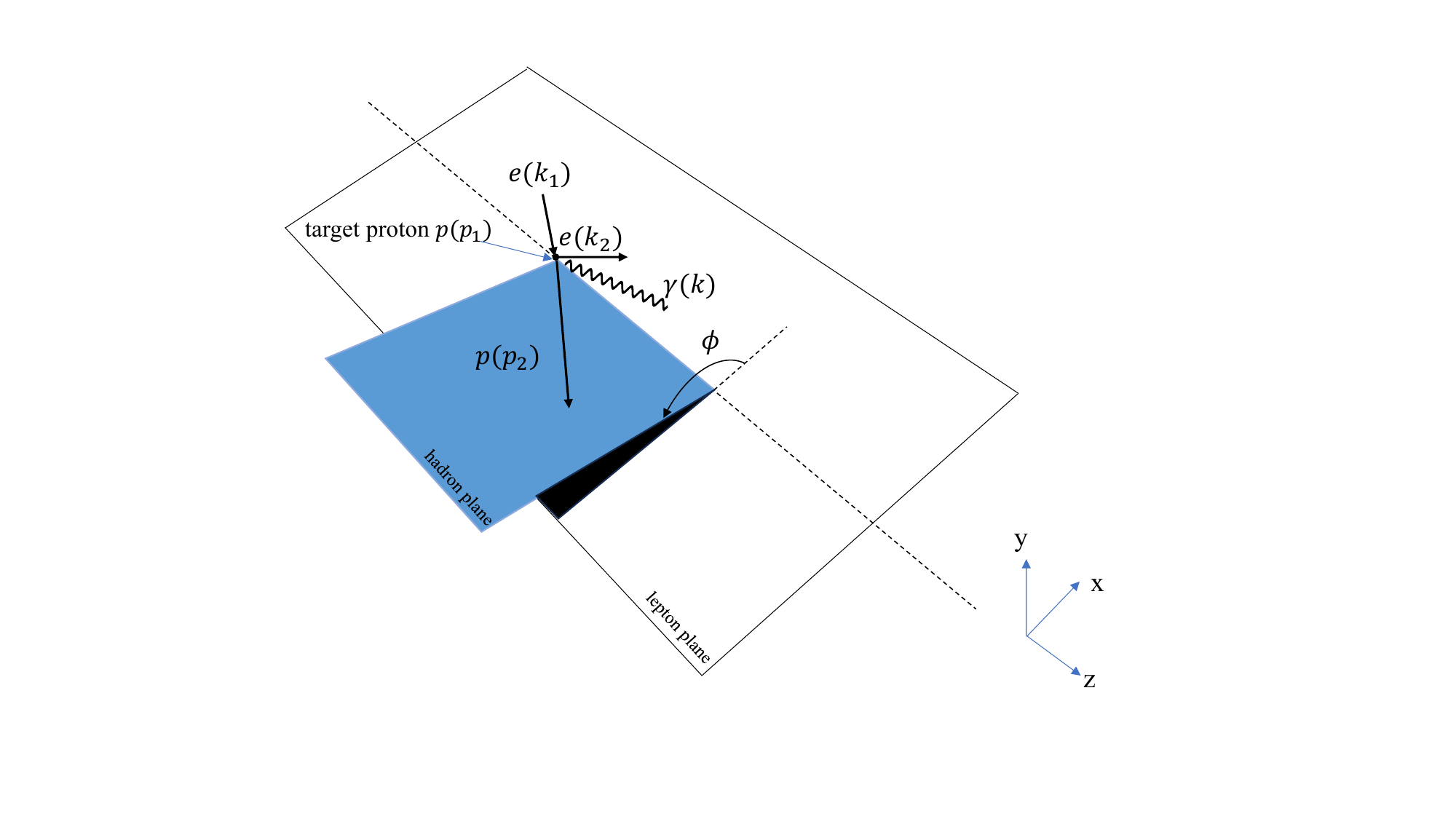}
    \caption{Definition of azimuthal angles for the $ep\to ep \gamma$ process in the target rest frame.}
    \label{fig:axes}
\end{figure}

The differential scattering cross-section is a crucial quantity that directly links theoretical calculations to experimental measurements. Due to the rotational symmetry of the scattered lepton around the incident axis in the laboratory frame, the scattering result is independent of the azimuthal angle $\phi_e$ of the lepton. By integrating over $\phi_e$,  
we obtain the following four-fold differential cross-section \cite{defurne2015e00}:
\begin{align}
    \frac{d^4\sigma}{dQ^2 dx_B dt d\phi } = \frac{\alpha_{QED}^3}{8\pi(s_e-M^2)^2 x_B}\frac{1}{\sqrt{1+\varepsilon^2}}\frac{1}{e^6}|\mathcal{M}|^2,
\end{align}
where the kinematic variables are defined as
\begin{align}
    \varepsilon^2 = 4M^2 x_B^2/Q^2, \quad s_e=(k_1+p_1)^2.
\end{align}

To compare experimental results with theoretical calculations,  variables such as $s,s_1,s_2,t_1,t_2$ are converted into functions of $Q^2,x_B,t,\phi$ and $E_k$, where $E_k$ is the energy of the incident lepton in the target rest frame:
\begin{align}
s =&m_e^2 + m_N^2 + 2m_N E_k,\\
s_1=&-\frac{-Q^2-m_N^2 x_B+Q^2 x_B}{x_B},\\
t_1=&-\frac{Q^2-m_N^2 x_B+t x_B}{x_B},\\
t_2=&t.
\end{align}
{From the equations above, we can derive a one-to-one correspondence between the variables $\{s, s_{1}, t_{1}, t_2\}$ and the kinematic quantities $\{Q^2, x_B, t, E_k\}$. Furthermore, by incorporating the expression for $\cos\phi$—which depends on the five Mandelstam variables—the value of $s_2$ can likewise be uniquely determined as a function of $Q^2, x_B, t, E_k,$ and $\phi$.}

\section{Comparison with Experimental Data and LECs Fitting}

The experimental data utilized in this analysis are obtained from the Hall A experiment at Jefferson Lab, as summarized in Tables VII, VIII, XII, and XIII of Ref.~\cite{defurne2015e00}. Each table corresponds to a distinct kinematic configuration. The specific kinematic ranges for each configuration are:

\begin{itemize}
    \item \textbf{Table VII:} $x_B = 0.34\text{--}0.38$, $Q^2 = 1.8\text{--}2.0\,\text{GeV}^2$, and $-t = 0.17\text{--}0.37\,\text{GeV}^2$;
    \item \textbf{Table VIII:} $x_B = 0.34\text{--}0.37$, $Q^2 = 2.2\text{--}2.4\,\text{GeV}^2$, and $-t = 0.17\text{--}0.37\,\text{GeV}^2$;
    \item \textbf{Table XII:} $x_B = 0.38\text{--}0.40$, $Q^2 = 2.0\text{--}2.1\,\text{GeV}^2$, and $-t = 0.19\text{--}0.37\,\text{GeV}^2$;
    \item \textbf{Table XIII:} $x_B = 0.33\text{--}0.34$, $Q^2 = 2.1\text{--}2.2\,\text{GeV}^2$, and $-t = 0.17\text{--}0.37\,\text{GeV}^2$.
\end{itemize}
In the first strategy,  we try to determine 
the low-energy constants  by fitting  the experimental datasets. While the fit quality at $\mathcal{O}(p^2)$ remains unsatisfactory due to the single adjustable parameter $c_6 + c_7$, the inclusion of $\mathcal{O}(p^3)$ corrections---  introducing the additional combination $d_6 + 2d_7$---yields a substantial improvement. The resulting global fit and the extracted parameters are presented in Figure~\ref{fig:tableAll_fit} and Table~\ref{tableAll_results}, respectively.
\begin{figure}[H]
    \centering
    \includegraphics[width=\textwidth]{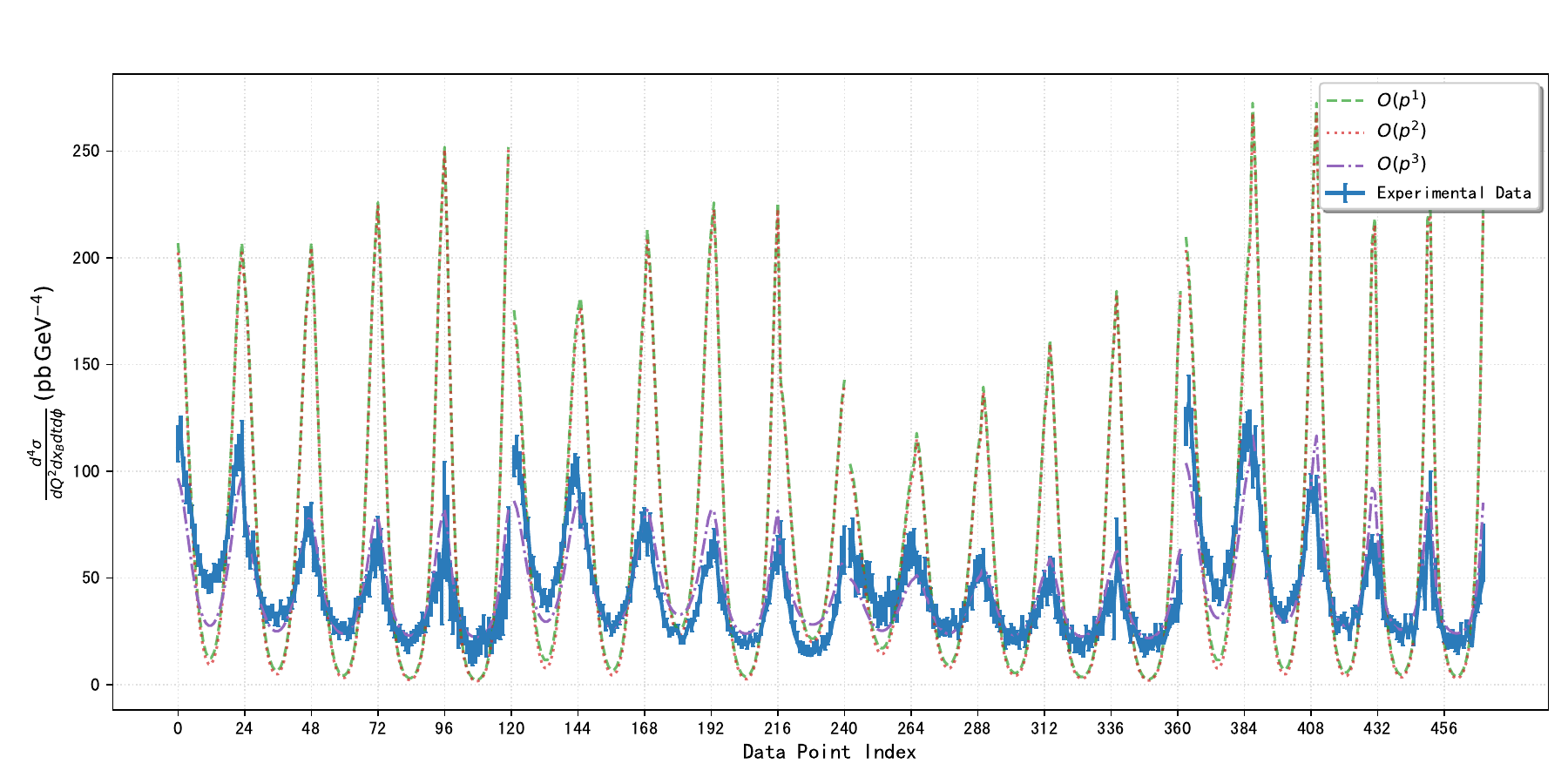} 
    \caption{Global fit results for all data sets from \cite{defurne2015e00}.The data are displayed sequentially according to Tables VII, VIII, XII, and XIII of Ref.~\cite{defurne2015e00}, reading the columns of each table from left to right.}
    \label{fig:tableAll_fit}
\end{figure}

\begin{table}[htbp]
\centering
\caption{Global fitted parameters $\alpha_1=c_6+c_7$ and $\alpha_2=d_6+2d_7$, and $\chi^2/dof$ values for different orders using the full dataset from \cite{defurne2015e00}.}
\label{tableAll_results}
\begin{tabular}{l c c}
\toprule
& \textbf{$O(p^2)$ } & \textbf{$O(p^3)$ } \\
\midrule
$\begin{array}{@{}l@{}}
\alpha_1 = c_6 + c_7, \\
\alpha_2 = d_6 + 2d_7
\end{array}$ 
& 
$\alpha_1 = -0.2786 \pm 0.0368$ 
& 
$\begin{aligned}
\alpha_1 &= 0.2114 \pm 0.0090, \\
\alpha_2 &= -1.1951 \pm 0.0098
\end{aligned}$ \\
\midrule
$\chi^2/\mathrm{dof}$ & 49.35 & 3.65 \\
\bottomrule
\end{tabular}
\end{table}

The global fit to the combined experimental dataset yields the parameter combinations $c_6+c_7=0.2114$ and $d_6+2d_7=-1.1951$. A comparative analysis reveals a significant discrepancy between these extracted values and those reported in Refs.~\cite{guerrero2020threshold,fuchs2004electromagnetic}, where the corresponding parameters after conversion of convention are approximately $c_6+c_7 \approx 2.39$ and $d_6+2d_7 \approx -1.68$.
 We attribute this discrepancy primarily to the kinematic region of the experiment. It is important to note that the variables $t$ and $Q^2$ characterize the virtualities (squared four-momentum transfers) of the photons associated with the BH and VCS processes, respectively. While the momentum transfer to the nucleon in the BH process remains moderate ($-t \in [0.17, 0.37] \text{ GeV}^2$), the virtual photon mass squared in the VCS process, $Q^2 \in [1.8, 2.4] \text{ GeV}^2$, is relatively large. This high-$Q^2$ regime    exceeds the  validity domain of standard Chiral Perturbation Theory ($\chi$PT), necessitating the {inclusion of loop contributions,  resonance contributions (most notably the $\Delta(1232)$ resonance) or $\rho$ meson effects to ensure a reliable theoretical description.} We leave this for future investigations. So, in the following analysis of low energy $lp$ scatterings, we still use the standard parameters from~Refs.~\cite{guerrero2020threshold,fuchs2004electromagnetic}.

\section{Low-Energy $l p$ Scattering}

As previously discussed, modern experimental precision has reached a level where the emission of photons can no longer be adequately addressed using the conventional soft-photon approximation. While low-energy elastic $lp$ scattering, including the two-photon exchange (TPE) box-diagram corrections, has been rigorously calculated within the framework of manifestly Lorentz-invariant chiral perturbation theory~\cite{PhysRevD.105.094008}, a complete description of the process requires a more detailed treatment of the radiative tail. 

Therefore, in this work, we aim to further investigate the differential cross-section for the emission of hard photons. Furthermore, in the context of ongoing $\mu p$ scattering experiments, the lepton mass $m_l$ can no longer be neglected as it significantly suppresses the phase space and modifies the radiation pattern. Without invoking any soft-photon approximations, the exact differential cross-section in the laboratory frame is given by~\cite{Gramolin_2014}:
\begin{align}
    \frac{\mathrm{d}\sigma}{\mathrm{d}E_{\gamma} \mathrm{d}\Omega_{\gamma} \mathrm{d}\Omega_{k_2}} 
    &= \frac{1}{(4\pi)^5} \frac{1}{m_N |\vec{k}_1|} 
    \sum_{E_{k_2}} \frac{E_\gamma |\vec{k}_2|^2 |\mathcal{M}|^2}{\left| A^{\prime} E_{k_2} - B^{\prime} |\vec{k}_2| \right|},
    \label{eq:diff_cross_section}
\end{align}
where $E_{\gamma}$ is the photon energy, $\Omega_{\gamma}$ is the solid angle of the emitted photon (characterized by the polar angle $\theta_{\gamma}$ with respect to the $z$-axis and the azimuthal angle $\phi_{\gamma}$), and $\Omega_{k_2}$ is the solid angle of the scattered lepton(same as $\Omega_{\gamma}$). The kinematic variables $A^{\prime}$ and $B^{\prime}$ are defined as follows:
\begin{align}
    A^{\prime} &= |\vec{k}_1|{\rm cos}\theta_{k_2}-E_{\gamma}{\rm cos}\psi, \\
    {\rm cos}\psi &= {\rm cos}\theta_{k_2} {\rm cos}\theta_{\gamma}+{\rm sin} \theta_{k_2} {\rm sin }\theta_{\gamma} {\rm cos}\phi_{\gamma},\\
    B^{\prime} &= E_1 + m_N - E_{\gamma},
\end{align}
As previously discussed, the variables $t$ and $Q^2$ serve as the critical kinematic parameters in the evaluation of radiative corrections. In the laboratory frame, they exhibit a dependence on the incident lepton energy $E_{\rm beam}$, as well as the final-state kinematic parameters: $E_{\gamma}, \theta_{\gamma}, \phi_{\gamma},$ and the scattered lepton angle $\theta_{k_2}$. 

With the specific objective of extracting the bremsstrahlung radiative correction background in low-$Q^2$ $\mu p$ scattering for the upcoming MUSE experiment, we impose constraints to restrict both $t$ and $Q^2$ to values below $0.08 \, \text{GeV}^2$. To this end, we perform a kinematic analysis selecting the lepton to be a muon ($m_l=0.1057\,\text{GeV}$).

It is self-evident that the incident energy dictates the kinematic constraints on $t$ and $Q^2$; specifically, lower incident energies restrict these variables to smaller   magnitudes. Beyond this energy dependence, the coupled correlation between the lepton and photon scattering angles plays a decisive role in determining the momentum transfer, the impact of which is summarized in the contour plots of Figure~\ref{fig:contour_scan}.

\begin{figure}[H]
    \centering
    \includegraphics[width=0.9\linewidth]{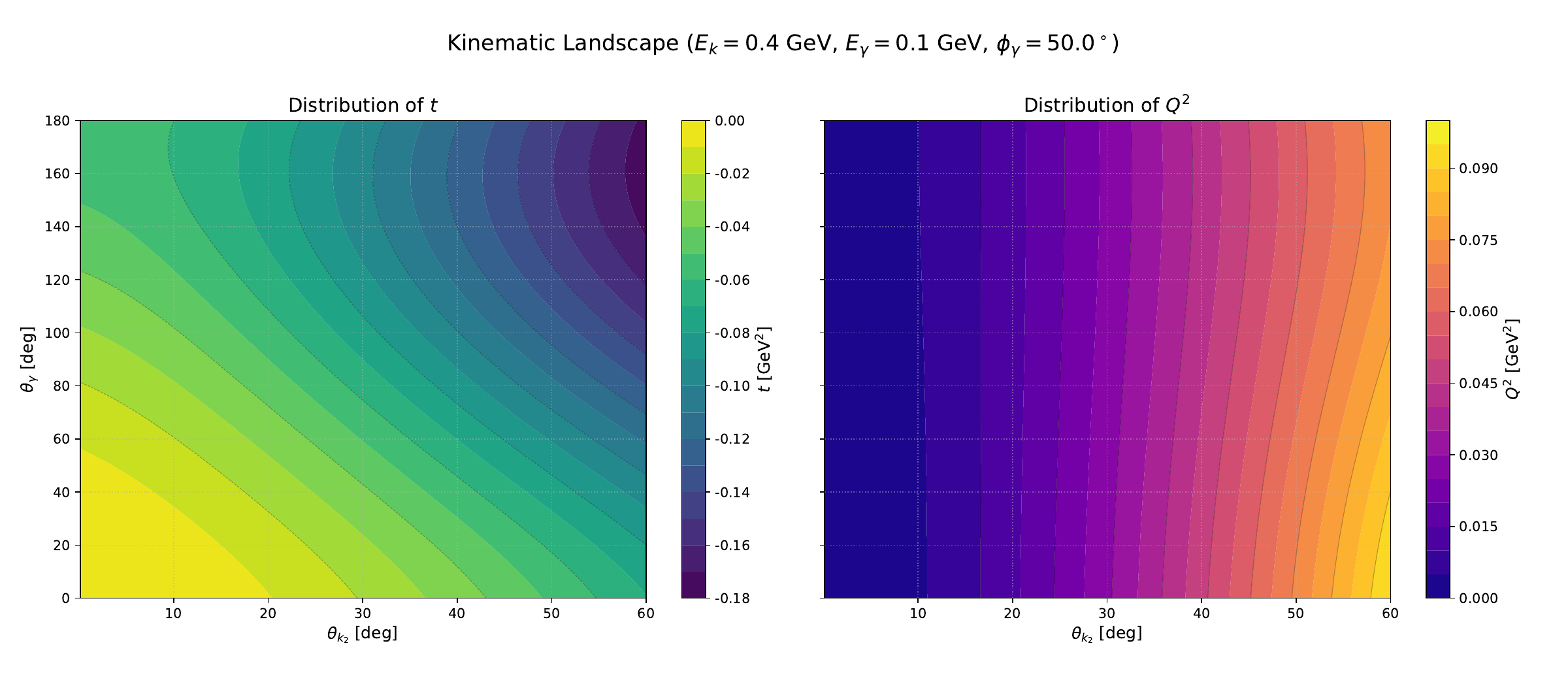} 
    \caption{Contour plots displaying the kinematic dependence of the momentum transfer $t$ (left panel) and the virtuality $Q^2$ (right panel) on the scattered lepton polar angle $\theta_{k_2}$ and the photon polar angle $\theta_\gamma$. The calculations are performed at an incident beam energy of $E_{\text{beam}} = 0.4\,\text{GeV}$ with a fixed photon energy $E_\gamma = 0.1\,\text{GeV}$ and azimuthal angle $\phi_\gamma = 50^\circ$.}
    \label{fig:contour_scan}
\end{figure}
Notably, the coupled influence of $\theta_{k_2}$ and $\theta_{\gamma}$ on the momentum transfer $t$ remains significant even at lower incident energies. Taking for examples, we present our theoretical predictions for the differential cross sections at a fixed photon scattering angle of $\theta_{\gamma} = 40^\circ$ and specific lepton angles $\theta_{k_2} = 15^\circ, 20^\circ$, and $25^\circ$, respectively, in Figs.~\ref{fig:15hard_photon_cross_section}--\ref{fig:25hard_photon_cross_section}. From a theoretical perspective, we suggest that experimental measurements prioritize a lepton scattering angle of $\theta_{k_2} < 30^\circ$. Such a kinematic selection helps to ensure the validity of the low-energy expansion and may provide a clearer observation of the targeted radiative effects.

Using the formula derived above, we present a theoretical prediction for the {`hard photon'} differential cross-section, as illustrated in ~\Cref{fig:15hard_photon_cross_section,fig:20hard_photon_cross_section,fig:25hard_photon_cross_section}. For this calculation, the low-energy constants (LECs) are adopted from Ref.~\cite{guerrero2020threshold}.

\begin{figure}[htbp]
    \centering
    \includegraphics[width=0.9\linewidth]{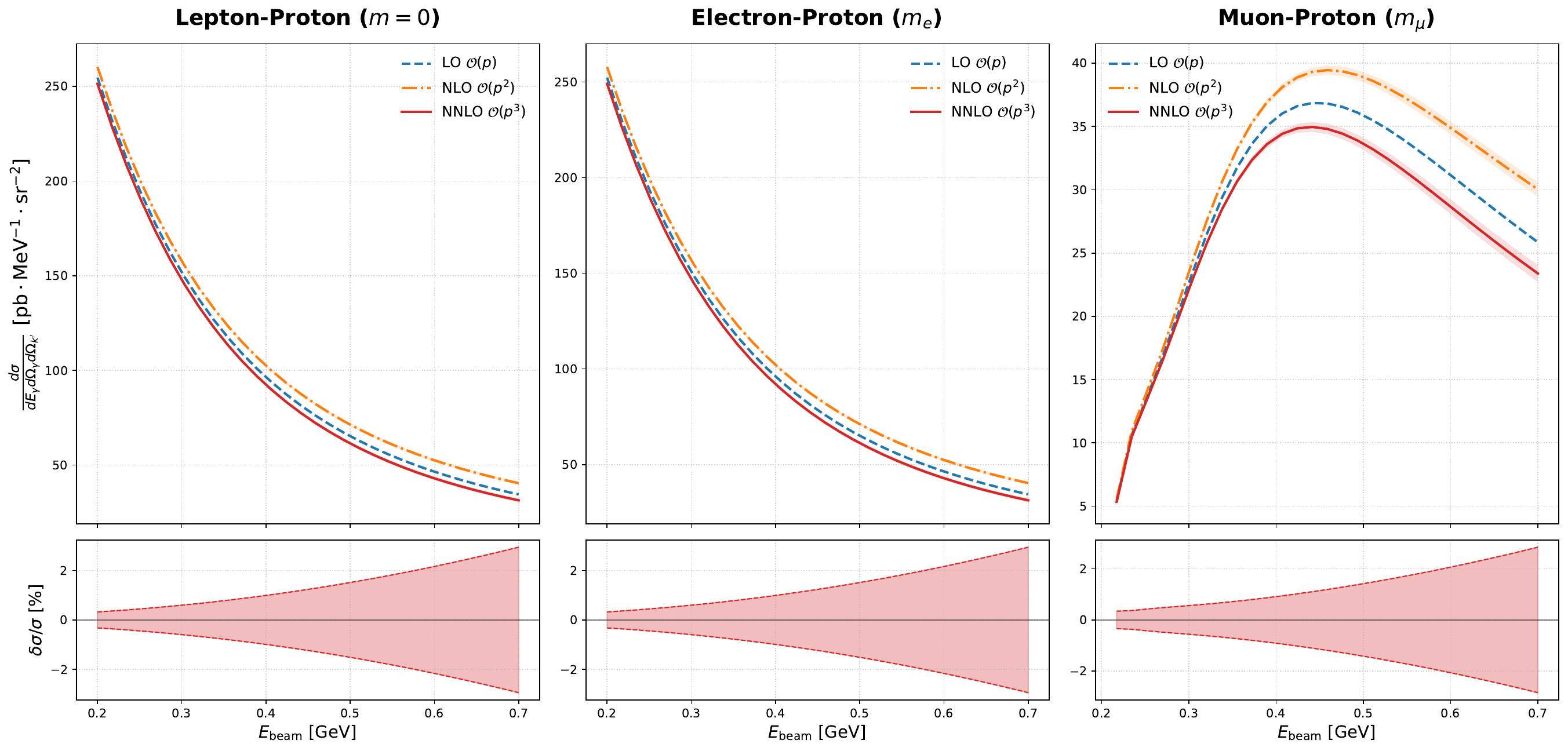} 
    \caption{Theoretical prediction of the differential cross-section for hard photon emission. The kinematic parameters are fixed at $E_\gamma = 0.1~\mathrm{GeV}$, $\theta_{k_2} = 15^\circ$, $\theta_\gamma = 40^\circ$, and $\phi_\gamma = 50^\circ$. The evolution of the momentum transfer variables with the beam energy $E$ is considered, covering the ranges $t \in [-0.042, -0.010]~\mathrm{GeV}^2$ and $Q^2 \in [0.006, 0.027]~\mathrm{GeV}^2$. The   low-energy constants (LECs) are taken from {Ref.~\cite{guerrero2020threshold}}.}
    \label{fig:15hard_photon_cross_section}
\end{figure}

\begin{figure}[htbp]
    \centering
    \includegraphics[width=0.9\linewidth]{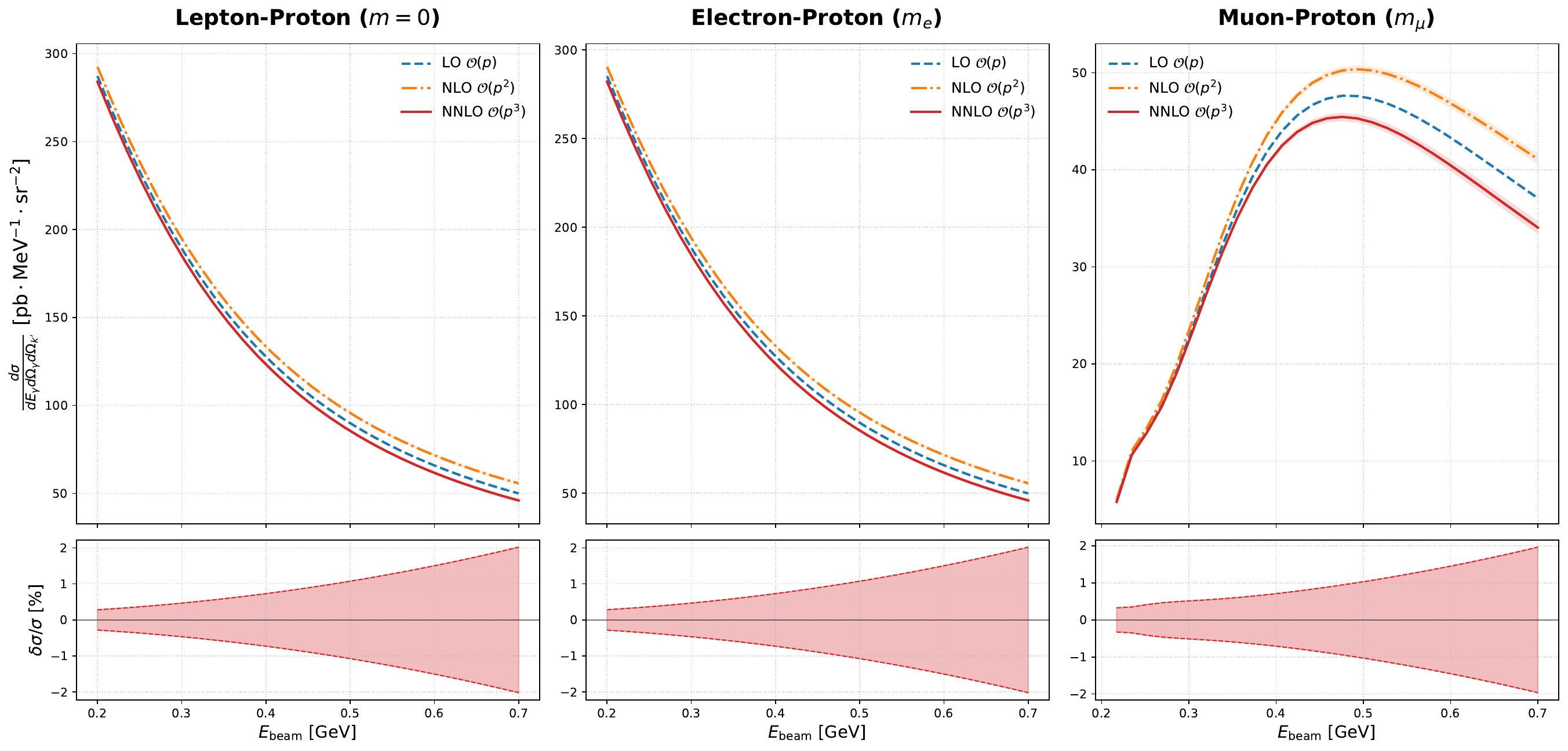} 
    \caption{Theoretical prediction of the differential cross-section for hard photon emission. The kinematic parameters are fixed at $E_\gamma = 0.1~\mathrm{GeV}$, $\theta_{k_2} = 20^\circ$, $\theta_\gamma = 40^\circ$, and $\phi_\gamma = 50^\circ$. The evolution of the momentum transfer variables with the beam energy $E$ is considered, covering the ranges $t \in [-0.063, -0.011]~\mathrm{GeV}^2$ and $Q^2 \in [0.008, 0.047]~\mathrm{GeV}^2$. The   LECs are taken from Ref.~\cite{guerrero2020threshold}.}
    \label{fig:20hard_photon_cross_section}
\end{figure}

\begin{figure}[htbp]
    \centering
    \includegraphics[width=0.9\linewidth]{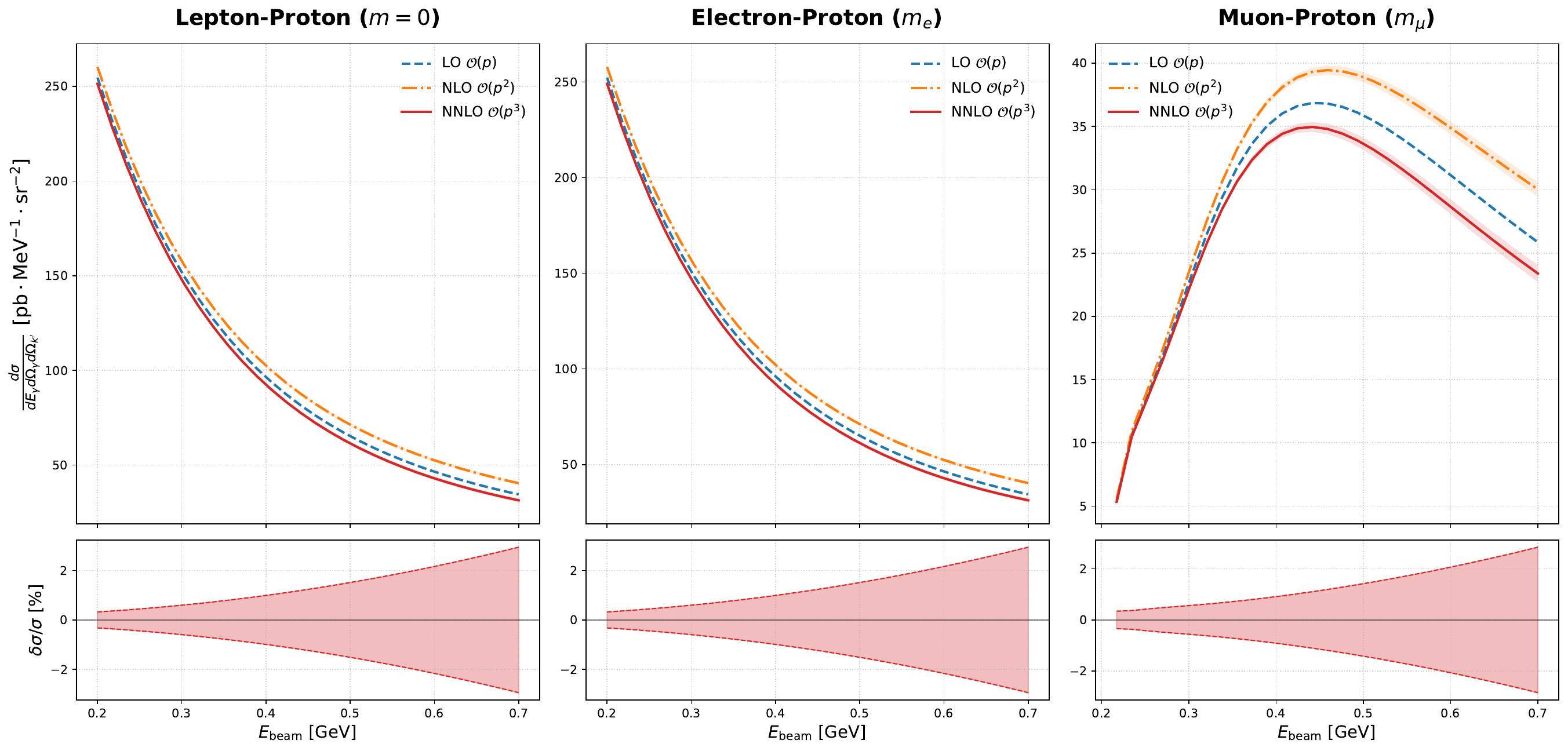} 
    \caption{Theoretical prediction of the differential cross-section for hard photon emission. The kinematic parameters are fixed at $E_\gamma = 0.1~\mathrm{GeV}$, $\theta_{k_2} = 25^\circ$, $\theta_\gamma = 40^\circ$, and $\phi_\gamma = 50^\circ$. The evolution of the momentum transfer variables with the beam energy $E$ is considered, covering the ranges $t \in [-0.089, -0.012]~\mathrm{GeV}^2$ and $Q^2 \in [0.010, 0.071]~\mathrm{GeV}^2$. The  LECs are taken from Ref.~\cite{guerrero2020threshold}.}
    \label{fig:25hard_photon_cross_section}
\end{figure}
From Figure~\ref{fig:15hard_photon_cross_section}--\ref{fig:25hard_photon_cross_section}, we observe that the impact of the electron mass on the results is negligible. However, for $\mu p$ scattering in the low-energy regime, there are significant modifications to the differential cross-sections, as seen by comparing~\Cref{fig:15hard_photon_cross_section,fig:20hard_photon_cross_section,fig:25hard_photon_cross_section}. Notably, the differential cross-section for $\mu p$ scattering exhibits two distinct features in this kinematic regime. First, the magnitude of the cross-section is numerically suppressed by approximately one order of magnitude compared to the electron case, reflecting the dynamical suppression induced by the heavier lepton mass. Second, in contrast to the monotonic behavior observed in the electron case, the $\mu p$ cross-section demonstrates a non-monotonic trend, initially increasing and subsequently decreasing with respect to the rising incident beam energy $E_{\rm beam}$. Furthermore, in this kinematic domain, the relative uncertainties of the differential cross-section $\delta\sigma/\sigma$ arising from the low-energy effective constants exhibit a noticeable energy dependence, expanding as the energy increases. Consequently, this  substantiates the necessity of the MUSE experiment and future low-energy scattering investigations.


\section{Conclusion}

In this work, we have performed a systematic calculation of the Bethe-Heitler (BH) and Virtual Compton Scattering (VCS) amplitudes for the $lp \to lp\gamma$ process within the framework of manifestly Lorentz-invariant $\chi$PT up to $\mathcal{O}(p^3)$, at tree level. In contrast to traditional radiative correction schemes that rely on the soft-photon approximation (SPA) or the ultra-relativistic limit, our approach retains the full kinematic dependence of the lepton mass and photon energy. This ensures the  validity of the framework in regimes involving massive leptons, such as muons, and the emission of `hard photons' { up to energies a few hundred MeV.}

A systematic analysis of the differential cross sections was conducted using the high-precision experimental data from JLab E00-110 \cite{defurne2015e00}. In the kinematic regime under consideration, the virtuality $Q^2$ exceeds the formal validity limit of the standard chiral expansion. 
{The necessity of considering the explicit $\Delta(1232)$ resonance and vector meson exchanges (such as the $\rho$ meson) in this high-momentum transfer region likely results in a significant shift of the LECs from their typical low-energy values.} We leave these subjects for future investigations. Also this process may provide a good opportunity to investigate photon coupling of the newly established subthreshold negative parity nucleon pole~\cite{Wang:2017agd,Cao:2022zhn}.

Furthermore, we investigated the impact of radiative corrections on future precision experiments, emphasizing the role of lepton mass in the low-energy regime. Our analysis demonstrates that at low energies, the coupled correlation between the lepton and photon scattering angles jointly dictates the validity domain of the chiral expansion. Notably, the non-zero lepton mass $m_\ell$ exerts a significant influence on the differential cross sections by modifying the kinematic phase space.  Our results can be compared with future experiments such as MUSE at PSI.


{\bf Acknowledgment:} We thank Xiong-Hui Cao at ITP, CAS, and D. L. Yao at Hunan University for helpful discussions. This work is supported by China National Natural Science Foundation under Contract
No. 12335002, 12375078. This work is also supported by ``the Fundamental Research Funds for the Central
Universities''.

\newpage

\bibliographystyle{utphys}
\bibliography{epref}

\end{document}